\documentclass[journal,twoside,web]{ieeecolor}
\usepackage{generic}
\usepackage{cite}
\usepackage{amsmath,amssymb,amsfonts}
\usepackage{algorithmic}
\usepackage{graphicx}
\usepackage{multirow}
\usepackage{textcomp}
\usepackage{tabularx}
\def\BibTeX{{\rm B\kern-.05em{\sc i\kern-.025em b}\kern-.08em
    T\kern-.1667em\lower.7ex\hbox{E}\kern-.125emX}}
\begin{document}

\title{BEATS: An Open-Source, High-Precision, Multi-Channel EEG Acquisition Tool System}

\author{Bing Zou, Yubo Zheng, Mu Shen, Yingying Luo, Lei Li, and Lin Zhang
\thanks{B. Zou, Y. Zheng, M. Shen, Y. Luo, L. Li, L. Zhang are with the School of Artificial Intelligence, Beijing University of Posts and Telecommunications, Beijing 100876, China (e-mail: leili@bupt.edu.cn).}}

\maketitle

\begin{abstract}
Stable and accurate electroencephalogram (EEG) signal acquisition is fundamental in non-invasive brain-computer interface (BCI) technology. Commonly used EEG acquisition systems' hardware and software are usually closed-source. Its inability to flexible expansion and secondary development is a major obstacle to real-time BCI research. This paper presents the Beijing University of Posts and Telecommunications EEG Acquisition Tool System named BEATS. It implements a comprehensive system from hardware to software, composed of the analog front end, microprocessor, and software platform. BEATS is capable of collecting 32-channel EEG signals at a guaranteed sampling rate of 4 kHz with wireless transmission. Compared to state-of-the-art systems used in many EEG fields, it displays a better sampling rate. Using techniques including direct memory access, first in first out, and timer, the precision and stability of the acquisition are ensured at the microsecond level. An evaluation is conducted during 24 hours of continuous acquisitions. There are no packet losses and the average maximum delay is only 0.07 s/h. Moreover, as an open-source system, BEATS provides detailed design files, and adopts a plug-in structure and easy-to-access materials, which makes it can be quickly reproduced. Schematics, source code, and other materials of BEATS are available at https://github.com/buptantEEG/BEATS.
\end{abstract}


\section{Introduction}
\label{sec:introduction}
\IEEEPARstart{E}{lectroencephalogram} (EEG) is a typical and potential non-invasive brain-computer interface (BCI) technology, which contains extensive information for clinical diagnosis and scientific research \cite{b1}. Recently, the application of EEG has expanded from medical health to intelligent control, workload detection \cite{b1}, sleep stage classification \cite{b2}, emotion recognition \cite{b3}, \cite{b4}, and other fields.

EEG acquisition is a primary and fundamental procedure of BCI \cite{b5}. At present, most EEG-related research depend on commercial EEG acquisition systems. Commercial systems can be quickly deployed and speed up the research progress. However, they have the following problems \cite{b6}, \cite{b7}. First, the hardware and software cannot be accessed publicly. Therefore, the EEG signals cannot be obtained in real time for online analysis. Researchers can only export data for offline analysis after an acquisition completes, which is a major obstacle for EEG-related research. Second, the system structure is fixed, and the iteration cycle is long. It is difficult to expand functions, and also lacks the ability to combine with the latest findings. Third, the commercial systems are expensive and bulky. The high price hinders large-scale applications and the huge size makes it difficult to deploy flexibly.

To alleviate the problems with commercial systems, some self-developed EEG acquisition systems have been proposed. Creamino \cite{b7} presents a cost-effective open-source EEG-based BCI system. It allocates 32-channel electrodes at a sampling rate of 500 Hz for a maximum speed of set up time, but still uses a wired communication method of universal serial bus (USB). \cite{b8} proposes a long-term acquisition system of forehead EEG, which implements 24-channel 250 Hz signal acquisition. In \cite{b9}, a modular board for EEG acquisition is developed and is able to capture 64 EEG channels at sampling rates up to 1 kHz and to transfer data over Bluetooth or Wi-Fi. \cite{b10} proposes a 32-channel EEG and electromyogram (EMG) acquisition system at a sampling rate of 2 kHz, which has a modular design and a good synchronization mechanism. \cite{b11} proposes a lightweight and affordable readout circuit with low-power, a 4 kHz sampling rate, and low-noise design for 8-channel EEG acquisitions named CochlEEG. \cite{b12} also increases the sampling rate to 4 kHz at 8-channel EEG acquisitions. Besides, there are some studies focused on electrodes \cite{b13}, \cite{b14}, and some focused on analog-to-digital conversion (ADC) and transmission of EEG signals \cite{b15}, \cite{b16}, \cite{b17}, \cite{b18}, \cite{b19}. Some of them also design a graphical user interface (GUI) for the signal processing and analysis \cite{b20}, \cite{b21}.

\begin{figure*}[tb]
    \centerline{\includegraphics[width=0.85\textwidth]{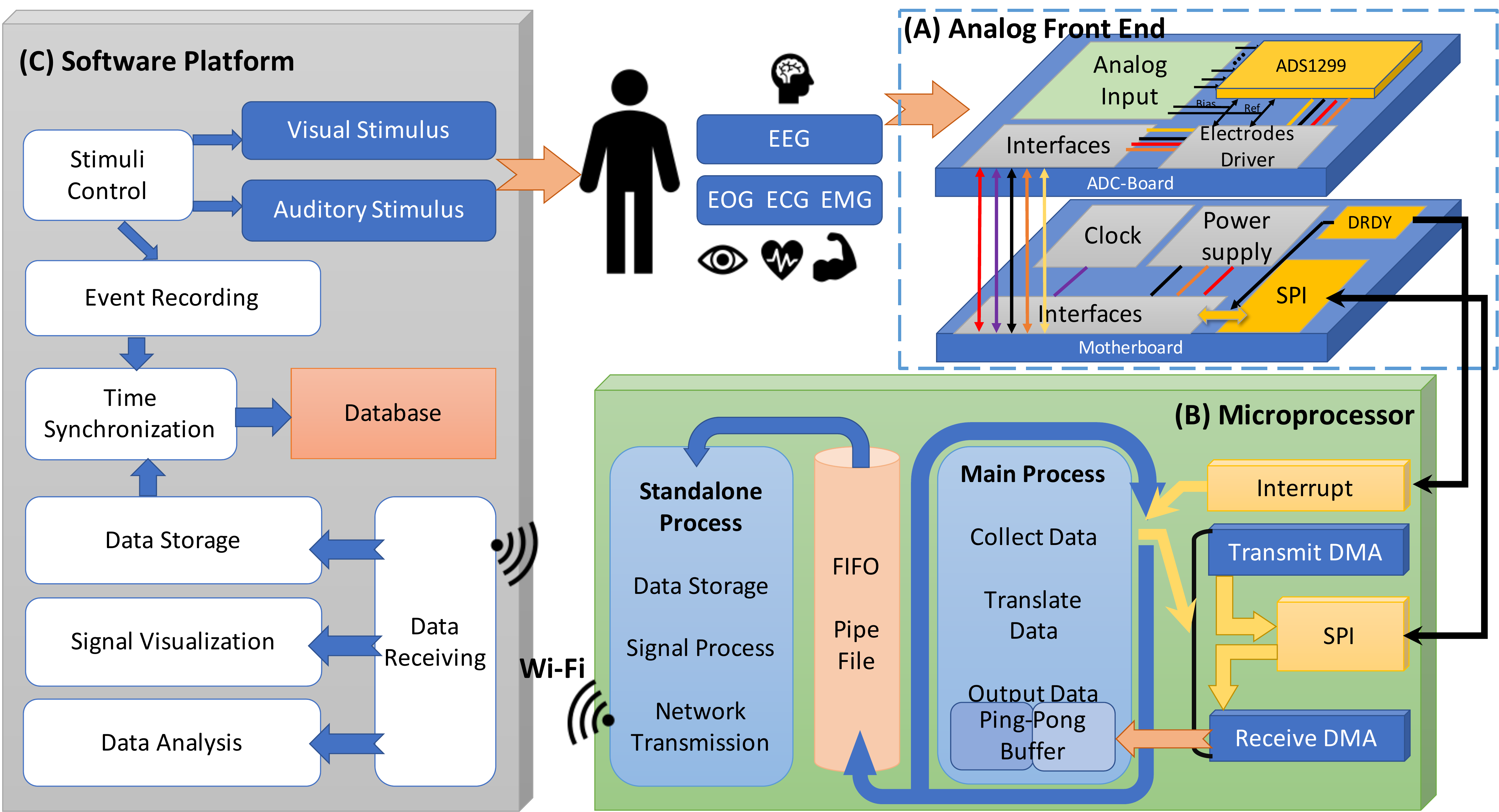}}
    \caption{The architecture of BEATS. Electrodes are firstly attached to human body according to the electrophysiological signals to be acquired. Then, these electrodes are connected to the AFE (A) to perform the analog-to-digital conversion. The converted signals are further processed by the microprocessor (B) and transmitted to the software platform (C). Meanwhile, the microprocessor also controls the process of acquisition and ensures stability and precision. After receiving the data transmitted from the microprocessor, the software platform visualizes and stores the signals in real time. Additionally, the software platform can generate visual and auditory stimuli to participants and record the corresponding time. After time synchronization between signals and experiments, the data with specific events can be stored for subsequent analysis.}
    \label{fig1}
\end{figure*}

However, most EEG acquisition products still share common shortcomings. First, The insufficient sampling rate (often 500 Hz or less) limits usage scenarios. EEG signals acquired under a higher sampling rate are generally considered to contain more detailed information \cite{b19}. Moreover, some studies can only be carried out under a high sampling rate, such as audiology applications (2 kHz) \cite{b11} and the detection of very high-frequency oscillations (VHFOs, over 1000 Hz) in epilepsy \cite{b22}. Second, the insufficient number of channels (under 5) leads to the loss of spatial information from different acquisition positions. For instance, EEG signals of different regions exhibit different characteristics used for different tasks in emotion detection \cite{b3}. Third, the connection between the analog front end (AFE) and signal processing back end still adopts wired communication methods, which cannot meet the needs of mobility. Although there are some studies that achieve good results on one metric, they often have downsides on another metric, such as the trade-off between the channel number and the sampling rate.

To address these shortcomings, this paper proposes the Beijing University of Posts and Telecommunications EEG Acquisition Tool System developed named BEATS, which is ideal for a wide range of BCI scenarios and long-term daily monitoring. The contribution of this work is:

\begin{enumerate}
    \item BEATS is capable of capturing 32 channels of EEG signals simultaneously at a guaranteed sampling rate of 4 kHz. Compared with most of the state-of-the-art EEG products used in many EEG fields \cite{b2}, \cite{b4}, \cite{b23}, it displays a better sampling rate.
    \item Using the underlying technology, a mechanism to ensure precision and stability in high-speed acquisition conditions is designed. After a 24-hour continuous evaluation, its average maximum delay is only 0.07 s/h, and there are no data loss or discontinuity.
    \item As an open-source system, it provides detailed design files and an instructional video, accompanied with printed circuit boards (PCBs) using easy-to-access materials and a plug-in structure, a user-friendly GUI and a wireless data transmission, which allows BEATS to be quickly and easily reproduced and set up.
\end{enumerate}

Schematics, source code, and other materials of BEATS are available at https://github.com/buptantEEG/BEATS. It mainly consists of three parts: the AFE, microprocessor (MP), and software platform (SW), as shown in Fig. \ref{fig1}. The AFE is conducted around an ADC chip to convert analog EEG signals to digital signals. The MP is applied to link the AFE with the SW, which takes charge of the configuration and control of the AFE and is used to transmit converted signals to the SW wirelessly. The SW implements many user-friendly operations, including signal visualization, data storage, stimuli generation and time synchronization, etc.

The rest of this paper is organized as follows. Section II, section III and section IV provide the details of the AFE, MP, and SW, respectively. Section V demonstrates system evaluations. Some application cases are introduced in section VI, and the conclusion is drawn in Section VII.

\section{Analog Front End}

The main function of the AFE is to acquire analog EEG signals and then convert them into digital signals. To make effective use of some common modules, two PCBs are designed as a plug-in structure, named the ADC-board and the motherboard. The ADC-board is designed to implement the data conversion, and the motherboard is designed to support the operation of the ADC-board. Multiple ADC-boards can be plugged into the same motherboard to achieve more channels, and the motherboard can be plugged into the MP to exchange data and commands.

\begin{figure*}[t]
	\centerline{\includegraphics[width=0.9\textwidth]{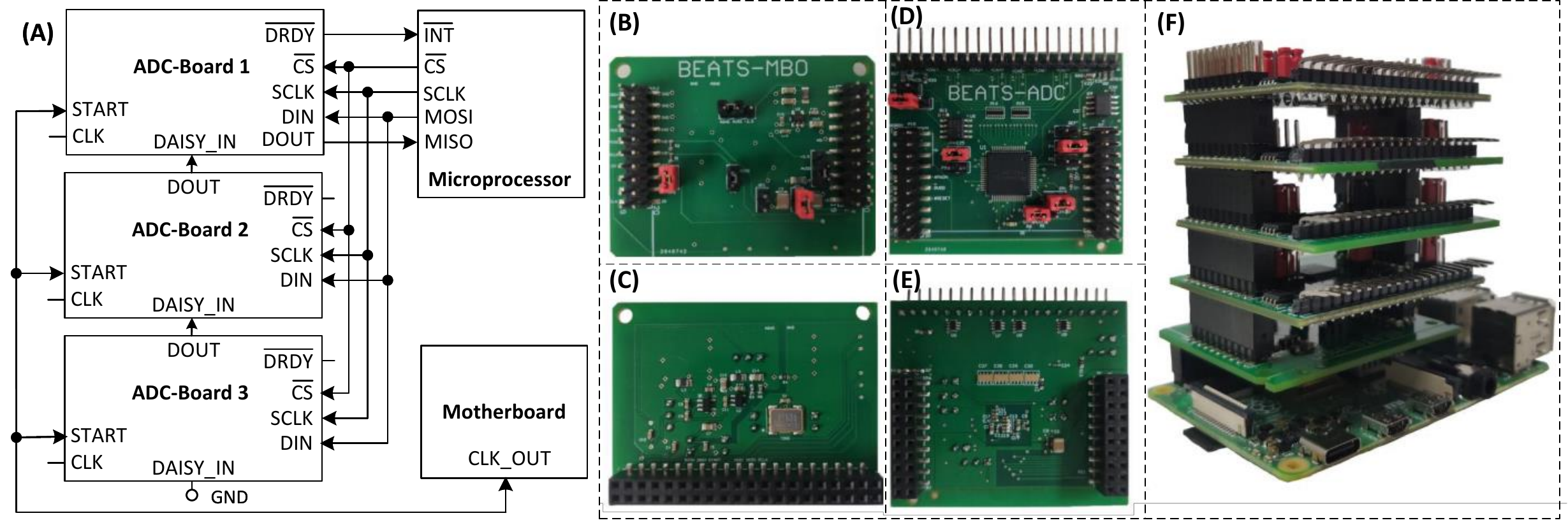}}
	\caption{(A) The scheme of 24-channel EEG acquisition in the daisy-chain mode. (B), (C) The front and back of the motherboard. (D), (E) The front and back of the ADC-board. (F) The picture of four ADC-boards, a motherboard and a MP plugged together.}
	\label{fig:pcb}
\end{figure*}

\subsection{ADC-Board Design}

\subsubsection{Analog-to-Digital Converter Module}

This module mainly carries out the conversion from analog signal to digital signal. The ADC chip used is ADS1299 developed by Texas Instruments (TI), which supports low-noise, 8-channel, 24-bit EEG measurements. To simplify the circuit to the greatest extent, we use as few pins as possible to implement the function of ADC. For some operations that can be done by pins or commands, we use command control instead of pin control, which can eliminate a lot of wiring.

In our design, only five signal pins are used to implement ADC operations, including reading conversion data, reading and writing registers, and controlling, etc. Four of the pins are used for serial peripheral interface (SPI) communication to exchange converted signals and control commands. The fifth used pin is a digital output pin named DRDY (Data Ready), which uses as a status signal to indicate when data are ready. DRDY goes low when new data are available, and it triggers the MP's response to fetch data.

\subsubsection{Analog Input Module}
When acquiring EEG signals, we use the SRB1 pin to route a common signal to multiple inputs for a referential montage configuration.
By connecting the reference electrode to the SRB1 pin and configuring the register, all positive input channels can share a common reference voltage as a negative input. In this way, only one common reference electrode is needed, and all the negative input electrodes can be saved.

Meanwhile, the positive and negative input pins of each channel are connected to a pin header for connecting differential input electrodes, corresponding to the INxN and INxP pins of the ADS1299. It can be used flexibly for a variety of connection scenarios.

For each analog input channel, an RC filter circuit is designed to suppress high-frequency interference. The resistance $R$ is 4.7 k$\Omega$ and the capacitance $C$ is 4700 pF, which constitutes a low-pass anti-aliasing filter. The cut-off frequency $F_{cutoff}$ is 7.2 kHz, which can meet the sampling requirements of 4 kHz, and can also meet the sampling requirements of up to about 16 kHz.
Additionally, anti-static components are added to avoid static damage to the circuit.

\subsubsection{Electrodes Driver Module}
In addition to the electrodes that collect signals, there is a reference electrode and a bias electrode. The reference electrode is connected to the SRB1 pin for a common reference. The bias electrode is used to sense the common-mode voltage of input electrodes and create a negative feedback loop by driving the body with an inverted common-mode signal. Then, the bias drive signal is generated by the feedback circuit and output to the human body through the bias electrode to suppress the common-mode noise, which is similar to the right leg drive of electrocardiogram (ECG) acquisition devices. In the schematic, some jumper pins are set up to conveniently switch the different functions of the reference and bias electrodes.

Two external operational amplifiers named OPA376 are used to buffer these electrodes and make the signal baseline in the center of the acquisition range. The first one is connected to the reference electrode to improve the driving ability of the reference signal. The other one is connected to the bias electrode to suppress the common-mode noise.

\subsection{Motherboard Design}

\subsubsection{Power Supply Module}
This module is designed to support the stable operation of the ADC chip, including a voltage inverter TPS60403, an adjustable voltage regulator TPS72325, and a high-accuracy voltage regulator with reverse current protection TPS73225. The power supply voltage is divided into analog high level (AVDD), analog low level (AVSS), and digital high level (DVDD), which are configured to 2.5 V, -2.5 V, and 3.3 V respectively. The original 5 V and 3.3 V voltages are supplied from the MP, and this module converts the 5 V voltage into $\pm$ 2.5 V. Meanwhile, the AVDD and AVSS can also be optionally set to 5 V and 0 V by jumpers, which should be used to avoid clipping when the input signal has a large positive DC offset. The ground pin of the chip is connected to the digital ground (DGND).

The system can be powered using a power bank or a wired connection. Using a common power bank with a rated capacity of 12500 mAh can support the system to work continuously for more than 24 hours. The rated voltage is 5 V and the power consumption is approximately 2.6 W per hour.

\subsubsection{Clock Module}
The built-in clock can support the basic operation of a single chip but is limited by the output and synchronization of multiple devices. To ensure clock consistency and provide a much more convenient way for multiple devices' simultaneous operations, the clock module is designed. It uses a high-precision clock chip named FXO-HC735 to provide a 2.048 MHz external clock. By connecting multiple devices to this module, the clock synchronization among multiple devices can be achieved.

\subsection{Multiple Devices Configuration}
A single ADC-board can achieve 8 channels of EEG acquisition, and multiple ADC-boards can be plugged together to expand the channel number. The scheme of 24-channel EEG acquisition is presented in Fig. \ref{fig:pcb} (A). There are two ways to connect multiple devices: cascade mode and daisy-chain mode. These modes typically need four signals: DIN, DOUT, SCLK, and CS. In the cascade mode, each device has its own CS signal. With one additional CS signal per device, multiple devices can be connected together. The number of signals needed to interface n devices is $3 + n$. By contrast, the daisy-chain mode reduces the SPI signals to four, regardless of the number of devices. In this work, the daisy-chain mode is applied to reduce and simplify connections.

The pin named DAISY\_IN is used for data transmission in the daisy chain mode. In this mode, SCLK, DIN, and CS signals are shared across multiple devices. The DOUT of the first device is connected to the MISO of the MP, and the DOUT of the second device is connected to the DAISY\_IN of the first device, thereby creating a chain. Data from the first device appear first on DOUT, followed by the data from the second device. If an additional ADC-board is needed, the DOUT of the latter device is connected to the DAISY\_IN pin of the former device. The DAISY\_IN pin is shorted to digital ground when not used. In addition, only the DRDY pin of the first board is connected with the MP, and the DRDY pins of other boards do not need to be connected. To ensure clock consistency, all PCBs use the same external clock from the motherboard.

\subsection{PCB Design and Manufacturing}
According to the schematic diagram, the PCBs are designed and manufactured with a plug-in structure to make setup more convenient. The pictures of PCBs are shown in Fig. \ref{fig:pcb}. The size of the ADC-board is 54.61 $mm$ $\times$ 44.20 $mm$, and the motherboard is 55.95 $mm$ $\times$ 43.38 $mm$. An ADC-board weighs 20 g, a motherboard weighs 15 g, and the MP weighs 45 g. The total weight of BEATS is 140 g when using four ADC-boards, a motherboard, and a MP for 32-channel EEG acquisition. 
After adding a 212 g battery, the weight is 352 g.

The motherboard adopts a 2-layer design. The top layer and bottom layer are signal layers, which are used to place components and wiring. The ADC-board has two more internal planes: the ground plane and the power plane. The copper at the circuit drawn by the plane mode is etched to divide it into several smaller planes to place different voltage sources. By dividing the internal plane, the device port and chip pins are connected directly with the corresponding network by via holes. The power plane is divided into three parts, corresponding to AVDD, AVSS, and DVDD, respectively. The ground plane layer is also divided into digital ground and analog ground. Digital ground and analog ground are connected by 0 $\Omega$ resistance to ensure the consistency of their potential. An etch line is added between the clock chip and the analog input to reduce the interference caused by clock oscillation.

\section{Microprocessor}

The MP is used to connect the AFE and the SW, as shown in Fig. \ref{fig1}. The MP used is Raspberry Pi (RPI) 4B which has rich peripheral interfaces and computing capabilities. Through SPI, the MP configures the programmable ADC chip and buffers the converted digital signals for packaging. Meanwhile, the MP wirelessly transmits the packaged data to the SW via Wi-Fi. Using underlying techniques, the embedded software is designed to ensure the operation precision and stability under a high sampling rate.

\begin{table*}[t]
\centering
\caption{Register Configurations}
\label{tab:reg}
\renewcommand\arraystretch{1.1}
\resizebox{\textwidth}{!}{%
\begin{tabular}{cc|c|c|c|c|c|c|c}
\hline
\multicolumn{2}{c|}{Address}                     & 0x01       & 0x02        & 0x03            & 0x05 - 0x0C & 0x0F        & 0x10        & 0x15   \\ \hline
\multicolumn{2}{c|}{Register}                    & CONFIG1    & CONFIG2     & CONFIG3         & CHxSET    & LOFF\_SENSP & LOFF\_SENSN & MISC1  \\ \hline
\multicolumn{1}{c|}{\multirow{9}{*}{Bits}} & Hex & 0x92       & 0xC0        & 0xEC            & 0x60      & 0x00        & 0x00        & 0x20   \\ \cline{2-9} 
\multicolumn{1}{c|}{}                      & 7   & 1          & 1           & PD\_REFBUF: 1   & PDx: 0    & LOFFP8: 0   & LOFFM8: 0   & 0      \\ \cline{2-9} 
\multicolumn{1}{c|}{} &
  6 &
  DAISY\_EN: 0 &
  1 &
  1 &
  \multirow{3}{*}{GAINx{[}2:0{]}: 110} &
  LOFFP7: 0 &
  LOFFM7: 0 &
  0 \\ \cline{2-5} \cline{7-9} 
\multicolumn{1}{c|}{}                      & 5   & CLK\_EN: 0 & 0           & 1               &           & LOFFP6: 0   & LOFFM6: 0   & SRB1:1 \\ \cline{2-5} \cline{7-9} 
\multicolumn{1}{c|}{}                      & 4   & 1          & INT\_CAL: 0 & BIAS\_MEAS: 0   &           & LOFFP5: 0   & LOFFM5: 0   & 0      \\ \cline{2-9} 
\multicolumn{1}{c|}{}                      & 3   & 0          & 0           & BIASREF\_INT: 1 & SRB2: 0   & LOFFP4: 0   & LOFFM4: 0   & 0      \\ \cline{2-9} 
\multicolumn{1}{c|}{} &
  2 &
  \multirow{3}{*}{DR{[}2:0{]}: 010} &
  CAL\_AMP0: 0 &
  PD\_BIAS: 1 &
  \multirow{3}{*}{MUXx{[}2:0{]}: 000} &
  LOFFP3: 0 &
  LOFFM3: 0 &
  0 \\ \cline{2-2} \cline{4-5} \cline{7-9} 
\multicolumn{1}{c|}{} &
  1 &
   &
  \multirow{2}{*}{CAL\_FREQ{[}1:0{]}: 00} &
  BIAS\_LOFF\_SENS &
   &
  LOFFP2: 0 &
  LOFFM2: 0 &
  0 \\ \cline{2-2} \cline{5-5} \cline{7-9} 
\multicolumn{1}{c|}{}                      & 0   &            &             & BIAS\_STAT      &           & LOFFP1: 0   & LOFFM1: 0   & 0      \\ \hline
\end{tabular}%
}
\end{table*}

\subsection{ADC Configuration}
Configuration information for the ADC is stored in its registers. By modifying the registers, the sampling rate, gain, working mode, and other settings can be configured. There is a read-only register ID at address 0x00, which is programmed during device manufacture and the lowest four bits should be 0xE if accessed correctly. Before configuring registers, the ID is used to determine whether the chip is working properly. Our register configurations are shown in Table \ref{tab:reg}.

In the CONFIG1, the DAISY\_EN is set to enable the daisy-chain mode for multiple devices' connection and the CLK\_EN is set to disable the clock output for allowing the use of external clock. The DR of the CONFIG1 is set to 010 for setting the sampling rate of 4 kHz. The CONFIG2 configures the test signal generation. With these bit settings, test signals are generated internally and are used to verify if the ADC is working properly. Through the CONFIG3, the bias driver is activated to reduce common-mode noise.

The CHnSET registers are used for 8-channel individual settings (n = 1 to 8). The GAINn is set to 110 for a maximum gain (24) of programmable amplifiers, which allows for better resolutions. By configuring the MUXn, the input channel is selected as the normal electrode input. In addition, the MUXn can also configure the input channels for testing internal noise and generating test signals. Through the SRB1 of the MISC1, the SRB1 pin is routed to all negative inputs as a common reference signal.

\subsection{Underlying Techniques}

During acquisition, the MP executes data fetching and transmitting tasks simultaneously. However, the central processing unit (CPU) of the MP will be distracted by other higher priority tasks, which leads a jitter in the time and makes the signal discontinuous. Therefore, underlying techniques are introduced to address these situations and ensure the stability and precision of the system, which are interrupt, first in first out (FIFO), and direct memory access (DMA).

\subsubsection{Interrupt}
During the signal acquisition, an interrupt is set to listen for the transitions of DRDY. Once DRDY goes low, the interrupt response is triggered. The CPU halts the action currently executed, instead performs the interrupt response to send the read data (RDATA) command and retrieve data immediately. After the interrupt response is complete, the CPU resumes the previous action. Since interrupt has the highest priority, the MP can retrieve data timely without causing data loss and time discontinuities.

\subsubsection{FIFO}
It is a communication pipe, whose data can be read and written by multiple processes independently at the same time. When the acquisition begins, a FIFO with a default size of 4096 bytes is established to buffer data. The data fetched from the AFE are put into the FIFO while the data to be transmitted to the SW are taken out from the other side of FIFO. The data firstly written into the FIFO will be read out first. This technique plays a major significance in the speed mismatch between data conversion and data transmission.

\subsubsection{DMA}
DMA is a memory access technique in RPI systems that allows the DMA controller to access memory independently of the CPU. We use DMA to move data from one memory address to another. In RPI, the SPI communications are configured by its registers. DMA can perform SPI communication by moving commands and values to the RPI's registers. This transmission action is initialized by the CPU, and completed by the DMA controller. The CPU can perform other tasks in the DMA data transmission process. Thus, the DMA controller is assigned to fetch data from the AFE through SPI regularly while the CPU can process data and transmit them to the SW in real time. 

\subsection{Embedded Software}

\begin{figure}[ht]
    \centering
    \includegraphics[width=0.43\textwidth]{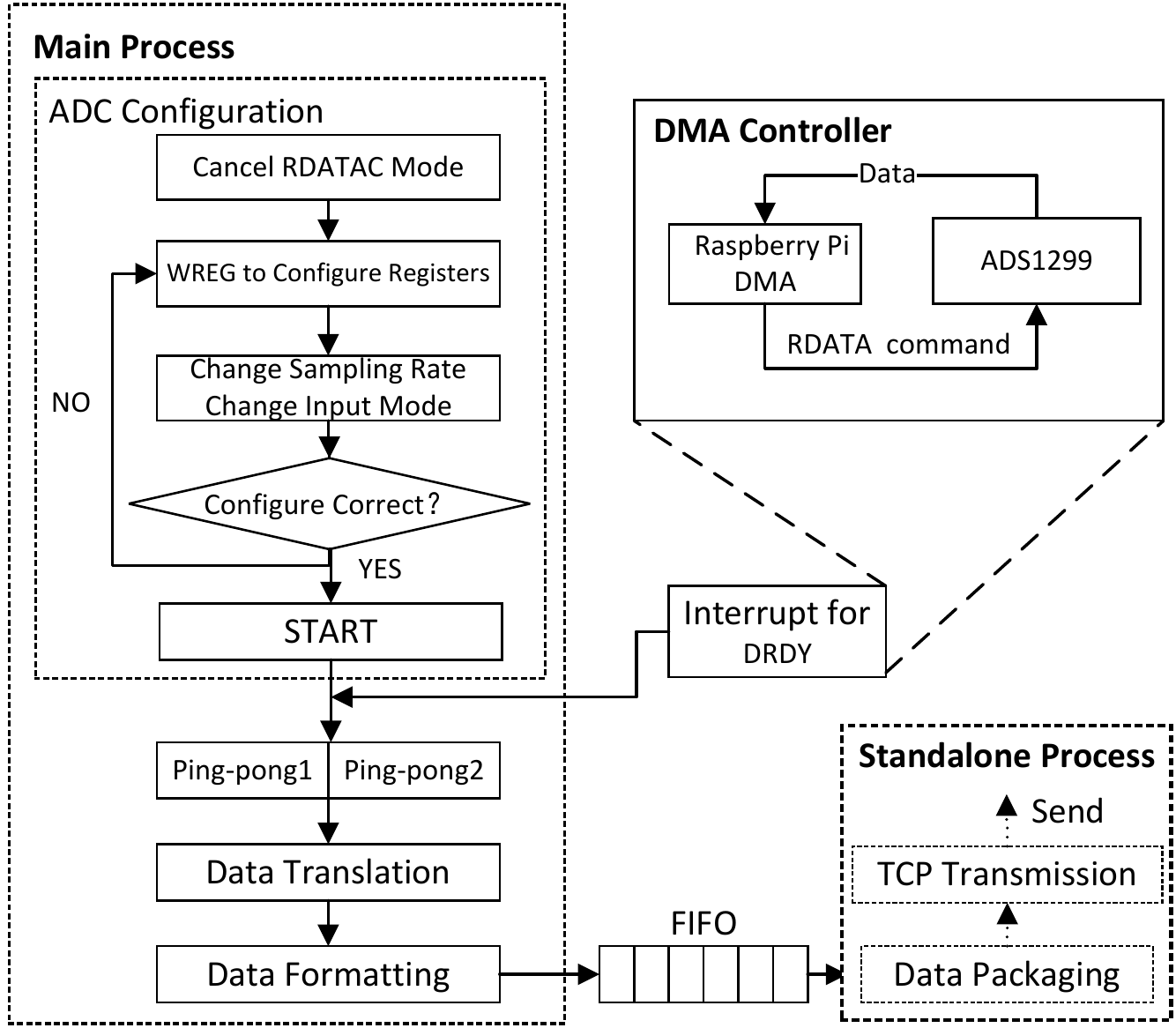}
    \caption{The flowchart of the Microprocessor embedded software.}
    \label{fig:RPI_flow}
\end{figure}

There are two processes in the embedded software shown in Fig. \ref{fig:RPI_flow}, named the main process and the standalone process. These two processes operate independently and will not jam each other. The ADC configuration is implemented first when the main process starts running, and its flow is as follows:

\begin{itemize}
    \item After power-up, the ADS1299 defaults to the read data continuous (RDATAC) mode. The SDATAC command is issued first to stop this mode and allow ADS1299 to accept and decode other commands.
    \item The registers are modified through the write to register (WREG) command and confirmed through the read from register (RREG) command.
    \item The START command is sent to enable data conversion.
    \item When detecting DRDY goes low, the RDATA command is used to fetch the latest converted signals, which provides a more stable and precise scheme to read data in a high-speed acquisition.
    \item As the acquisition is complete, the further data conversion is terminated by using the STOP command.
    \item If ADS1299 does not operate as expected, the RESET command is issued to set all registers to default values.
\end{itemize}

After the configuration, the interrupt and DMA controller are initialized for fetching data. The fetched data is cached in a ping-pong buffer first, which has two buffers that one buffer is being written while the other is being processed. These buffers work alternately so that data fetching and subsequent actions will not block each other. The buffer size is 40 samples, which works well in practice. The data fetched from ADS1299 is in binary twos complement format. The main process converts them into a decimal form, which calls data translation. Then, the decimal data are arranged in a certain format and outputted to FIFO finally. The main process executes the data translation and formatting continuously, and turns to data acquisition once the interrupt is triggered.

The data outputted to FIFO are further packaged and transmitted by the standalone process. In this process, the decimal data, timestamp, and channel status are taken out from FIFO and packaged in JavaScript Object Notation (JSON) packet. Each packet contains 160 samples. Then, a TCP connection is established between the MP and SW, and data are finally sent to the SW through Wi-Fi embedded in the RPI.

\begin{figure*}[htb]
    \centering
    \includegraphics[width=0.73\textwidth]{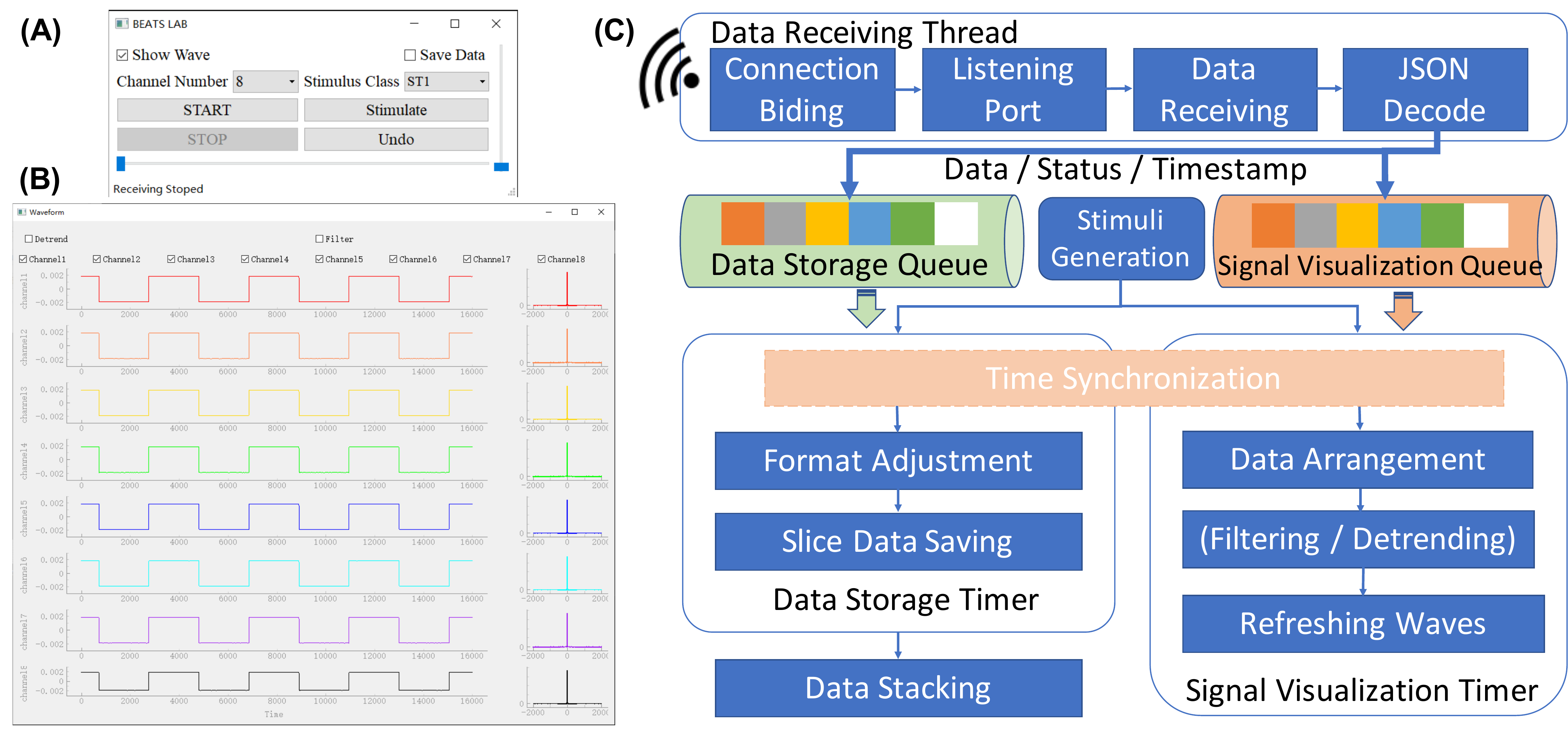}
    \caption{(A) The control window. (B) The waveform window. A test signal of single-frequency square wave is displayed, which indicates that the system is working properly. (C) The bottom architecture of the software platform.}
    \label{fig:gui}
\end{figure*}

\section{Software Platform}

To simplify users' operations, a GUI is designed to implement signal visualization, data storage and other interactions. It includes two windows, namely the control and the waveform window, as shown in Fig. \ref{fig:gui} (A) (B). At the bottom of the GUI, threads and timers are used to ensure real-time performance of the SW. In addition, the stimuli generation and time synchronization functions are developed to perform the alignment of signal and event timestamps during experiments. The software platform and the application cases in the next section are deployed on computers with the Windows 10 operating system. The computer used is equipped with an Intel Core i7-8550U CPU @ 1.80 GHz, 16 GB of RAM, and 512 GB of Hard Disk, and a similar or better computer configuration should work well.

\subsection{Graphical User Interface}
 
\subsubsection{Control Window}

This window is used to control and configure the acquisition process. The ``Save Data'' checkbox enables data storage function, whereas the ``Show Wave'' checkbox enables waveform display function and opens the waveform window. The vertical slider on the right and the horizontal slider on the bottom are used to adjust the amplitude and time range of the signals.

The ``Stimulate'' button is used to record the timestamp when it is pressed, and the ``Stimulus Class'' combobox is used to select the stimulus to be recorded. It provides a quick way to record some less time-critical stimuli. Meanwhile, the last stimulus recording can be revoked by the ``Undo'' button to avoid accidental presses.

\subsubsection{Waveform Window}
This window is used to display the time and frequency domain waveforms in real time. The channel number, ranges are determined in the control window. During acquisitions, the ``Filter'' checkbox is selected to remove the power frequency interference, and the ``Detrend'' checkbox is selected to remove baseline drift. The validation of test signal is presented in Fig. \ref{fig:gui} (B), which displays a single-frequency square wave and indicates that the system is working properly.

\subsection{Threads and Timers}
The bottom architecture of the SW is shown in Fig. \ref{fig:gui}, which including the data receiving thread, signal visualization timer, and data storage timer. The main process of the SW maintains the presentation of the GUI and the concurrent operations of threads and times.

\subsubsection{Data Receiving Thread}
This thread opens the port for transmission control protocol (TCP) connection first and listens to its status, so as to determine whether the MP has sent the data. When receiving data packets, it collects the packet length first, and then collects the packet according to the length, so as to solve the packet sticking problem of TCP transmission. Besides, a buffer is set manually to solve the packet breaking problem, which may destroy the independence of the data packet and affect the continuity of subsequent decoding. After receiving a complete JSON packet, the thread decodes the JSON string and puts them into the data queues. Through these queues, data can be shared with the timers.

\subsubsection{Signal Visualization Timer}
At the main process, this timer is set to trigger the refresh of the waveform window. When the time is up, the response function is performed to check the status of the signal visualization queue and update the waveform by fetching data from the queue. The time domain waveform will slide from back to front, and its refresh time is 1 ms. If the filtering or detrend function is enabled, they are also processed in this timer response. The use of timers can avoid the resource consumption caused by repeatedly querying queue status when the data is not ready. Besides, when the timer response is triggered, a new thread is generated to perform operations so that these operations will not block each other.

\subsubsection{Data Storage Timer}
Similar to the signal visualization timer, this time is set to save the data periodically and not affect the precise execution of the data receiving thread. When the time is up and the data storage queue is not empty, the timer response gets the data and stores them. The system input and output (IO) rate is usually slower than the network transmission rate. To avoid the delay caused by storing a large amount of data and solve the rate mismatch between network transmission and system IO, the data are divided into 5 ms segments for storage. At the end of the acquisition, the data are stacked into a complete file.

\subsubsection{Online Analysis Thread}
In addition to these basic functions, BEATS can also perform signal analysis at the same time as signal acquisition, which we call online analysis. To achieve this function, it is first necessary to create a queue to collect raw data for analysis, which is paralleling with the data storage queue and the signal visualization queue. Then, a new thread is created to pull data from this queue and use the data for real-time analysis. The statistical algorithm or trained model can be deployed in this thread.

\subsection{Stimuli Generation and Time Synchronization}
To observe the response of EEG signals to certain events, stimuli generation and time synchronization functions are developed. In the control window, timestamps of some basic stimuli are recorded when the button is pressed. Experimenters can apply immediate stimuli at any time based on their judgements. Meanwhile, an automated program is also developed to present visual and auditory stimuli, and to automatically record the timestamps of these stimuli. It can be used for some experimental data acquisition with predetermined steps.

The stimuli with timestamps are generated in the SW whereas the signals with timestamps are recorded in the MP, and they are synchronized through the alignment of timestamps. This timestamp is recorded in Coordinated Universal Time (UTC), and the majority of network devices support this standard.
The main process of alignment is as follows:
\begin{itemize}
    \item In the MP, data are continuously acquired, and the timestamp is also recorded for each data while acquiring data.
    \item In the SW, the moment of the stimulus generation is recorded by timestamp in UTC.
    \item The time-stamped data are transmitted to the SW continuously.
    \item After receiving and decoding the data from the MP, the SW  synchronizes the stimulus generation with the EEG acquisition based on the timestamps.
\end{itemize}

On both the MP and SW sides, the timestamp recording process is independent. Since the timestamps are recorded at the moment of stimulus and signal occurrence, the exact moment of the stimulus generation with the EEG acquisition can be synchronized. Because this scheme relies on the network transmission before it can be synchronized, there will be a few milliseconds delay in the results presentation. After several evaluations, this delay averaged 30 ms. For higher accuracy requirements, it can still be improved by specialized protocols \cite{b24} or hardware and we will further investigate these.

\section{System Evaluations}

\subsection{Clock Evaluation}
Under actual operating conditions, the internal clock is measured in the ADC-board, and the external clock is measured in the motherboard with the the ADC-board plugged. The results are shown in Table \ref{tab:clock}. The ADC chip accepts a clock range of 1.5 to 2.25 MHz. The internal and external clock all meet the requirements. In addition, the internal clock has an error rate of 3301 part per million (ppm) whereas the external clock has a much smaller error rate of 24 ppm.

\begin{table}[ht]
\centering
\caption{Clock Evaluation}
\label{tab:clock}
\begin{tabular}{ccc}
\hline
Clock     & Frequency   & Error Rate \\ \hline
Reference & 2.048 MHz   & 0 ppm      \\
Internal & 2.04165 MHz & 3101 ppm   \\
External & 2.04805 MHz & 24 ppm \\ \hline
\end{tabular}
\end{table}

\subsection{Noise Evaluation}
The input electrode of each channel is short-circuited to the reference electrode by configuring the MUXn of the CHnSET to 001. Then, the input short circuit noise of ADC-board is measured, including the root mean square of noise ${V_{RMS}}$, peak-to-peak value ${V_{PP}}$, effective number of bits (ENOB) and dynamic range. ENOB is calculated by (\ref{eq:enob}) and dynamic range is calculated by (\ref{eq:dr}). 

\begin{equation}
ENOB=log_{2}(\frac{V_{REF}}{\sqrt{2}\times Gain \times V_{RMS}}) \label{eq:enob}
\end{equation}
\begin{equation}
Dynamic\ Range = 20 \times log_{10}(\frac{V_{REF}}{\sqrt{2}\times Gain \times V_{RMS}}) \label{eq:dr}
\end{equation}

Evaluations are performed at sampling rates from 250 Hz to 4 kHz, and the results are shown in Table \ref{tab:noise}. The measured values are all close to the reference value, which reaches the acquisition requirements. The small drop in the measured value compared to the reference value is due to the extra noise introduced by the peripheral circuits included in the PCBs.

\begin{table}[ht]
\centering
\caption{The results of input short circuit noise \\ (Measured / Reference)}
\label{tab:noise}
\resizebox{\columnwidth}{!}{%
\begin{tabular}{ccccc}
\hline
\begin{tabular}[c]{@{}c@{}}Sampling\\ Rate (Hz)\end{tabular} &
  \begin{tabular}[c]{@{}c@{}}$V_{RMS}$\\ ($\mu{V}$)\end{tabular} &
  \begin{tabular}[c]{@{}c@{}}$V_{PP}$\\ ($\mu{V}$)\end{tabular} &
  ENOB &
  \begin{tabular}[c]{@{}c@{}}Dynamic\\ Range (dB)\end{tabular} \\ \hline
250  & 0.15 / 0.14 & 1.00 / 0.98 & 19.75 / 19.85 & 118.9 / 119.5 \\
500  & 0.20 / 0.20 & 1.54 / 1.39 & 19.34 / 19.35 & 116.4 / 116.5 \\
1000 & 0.28 / 0.28 & 2.21 / 1.97 & 18.85 / 18.85 & 113.5 / 113.5 \\
2000 & 0.40 / 0.40 & 2.97 / 2.79 & 18.34 / 18.35 & 110.4 / 110.4 \\
4000 & 0.57 / 0.56 & 4.51 / 3.94 & 17.83 / 17.84 & 107.3 / 107.4 \\ \hline
\end{tabular}
}
\end{table}

\subsection{Common-Mode Rejection Ratio (CMRR) Evaluation}
CMRR shows the suppression ability to the common-mode signal, which calculation method is shown in (\ref{eq:cmrr}). $A_{d}$ represents the voltage amplification factor of the differential mode signal and $A_{cm}$ represents the voltage amplification factor of the common mode signal. The differential mode signal is a 100 $\mu{V}$, 0 Hz to 70 Hz sine wave signal, and the common-mode signal is a 4.4 V, 0 Hz to 70 Hz sine signal with a DC bias of 2.5 V.

\begin{equation}
    CMRR = 10 \times log_{10}(\frac{A_{d}}{A_{cm}})^{2} = 20 \times log_{10}(\frac{A_{d}}{A_{cm}})
    \label{eq:cmrr}
\end{equation}

\begin{figure}[ht]
    \centering
    \includegraphics[width=0.48\textwidth]{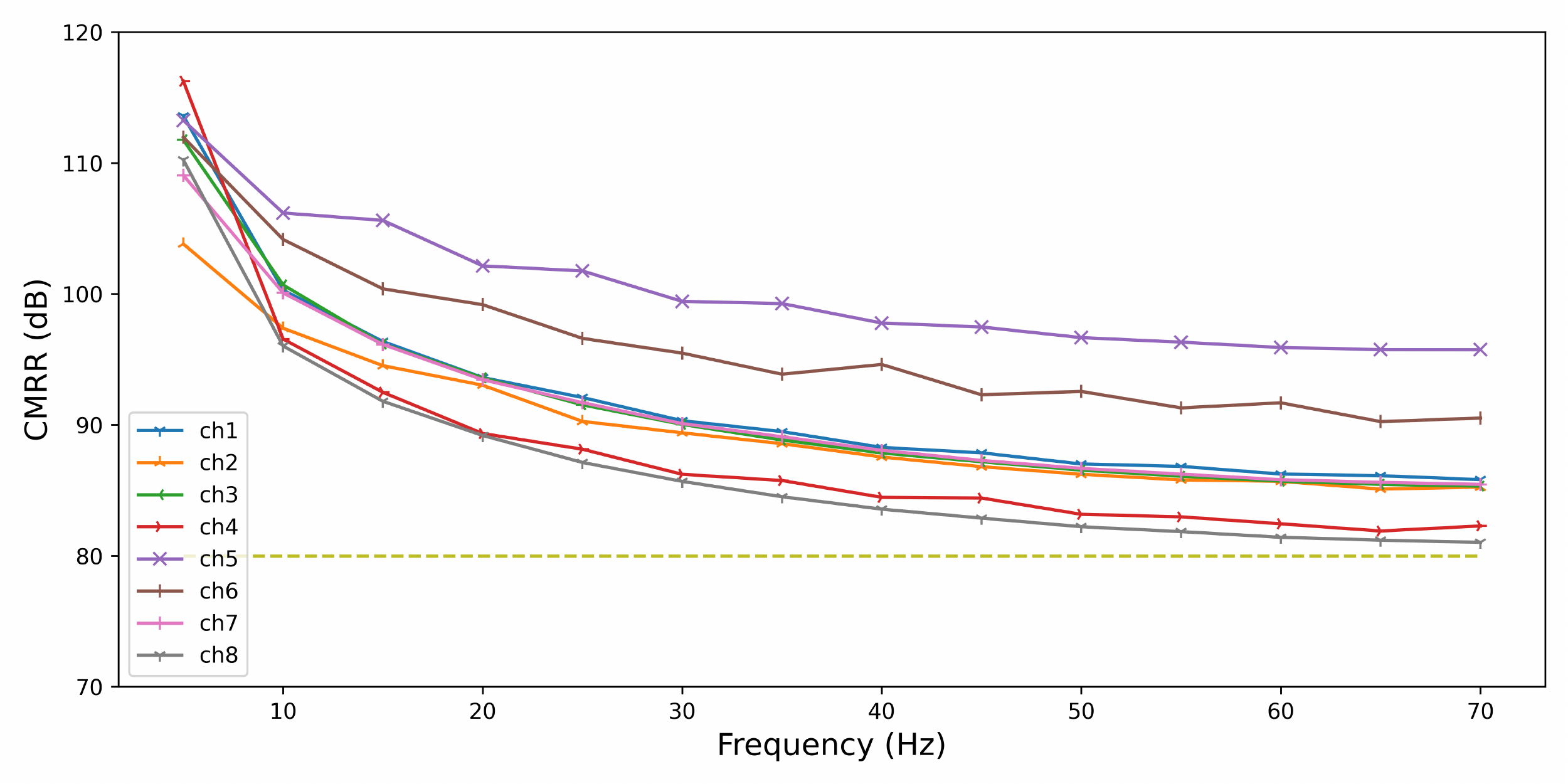}
    \caption{The results of CMRR in different frequencies.}
    \label{fig:cmrr}
\end{figure}

The CMRR of five ADC-boards from the same batch is evaluated and averaged. The results are shown in Fig. \ref{fig:cmrr}. CMRR decreases with the increase of frequency. The CMRR of each channel is higher than 80 dB, and meets the requirement of EEG equipment. The CMRR of different channels is slightly different. The possible reason is that CMRR is affected by circuit symmetry and the inevitable asymmetric layout of PCBs leads this difference. Additionally, there is a certain error range in the accuracy of the resistors, which also increases this difference.

\subsection{Delay and Loss Evaluation}

Delay and loss are used to evaluate the precision and stability of BEATS. The delay contains four dimensions. In the MP, the ``ADC'' represents the process of ADC and outputting data to FIFO, and the ``Trans'' represents the process of fetching data from FIFO and transmitting them to the SW. In the SW, the ``Save'' represents the process of data receiving and storage, and the ``Plot'' represents the process of data receiving and visualization on the GUI. The loss includes two dimensions, which reflects the data integrity. The ``MP'' represents the packet loss causing by data congestion and any other reason in the MP, whereas the ``SW'' represents the packet loss in the SW. The evaluations are conducted several times with different durations at the sampling rate of 4 kHz, the average results are shown in Table \ref{tab:delay}.

\begin{table}[ht]
\centering
\caption{Delay and Loss Evaluation}
\label{tab:delay}
\begin{tabular}{c|cccc|cc}
\hline
\multirow{2}{*}{\begin{tabular}[c]{@{}c@{}}Time\\ (hours)\end{tabular}} & \multicolumn{4}{c|}{Delay (s)} & \multicolumn{2}{c}{Loss (packets)} \\
    & ADC    & Trans           & Save            & Plot            & MP & SW \\ \hline
0.5 & 0.0874 & 0.1274          & \textbf{0.1595} & \textbf{0.1595} & 0   & 0  \\
1   & 0.6214 & \textbf{0.6369} & 0.4951          & 0.4951          & 0   & 0  \\
4   & 1.8049 & \textbf{1.8395} & 1.6305          & 1.6305          & 0   & 0  \\
8   & 1.4218 & \textbf{1.6232} & 0.6460          & 0.6460          & 0   & 0  \\
24  & 1.7311 & \textbf{1.7712} & 0.8198          & 0.8198          & 0   & 0  \\ \hline
\end{tabular}
\end{table}

At each acquisition duration, the maximum delay in the four dimensions is marked in bold. The delay does not increase with the increase of the acquisition duration, but is all in the range of about 0 to 2 s. It shows that BEATS can maintain a steady work when performing continuous data acquisition for at least 24 hours without gradually becoming stuttering, slower or blocked. There is a fixed delay over each duration, which may be accumulated from the system initialization to the stable operation. Even in continuous 24 hours of acquisition, the delay remains within the acceptable range, with an average maximum delay of 0.7 s/h. Meanwhile, the delay of the four dimensions does not differ much, and the delay of signal visualization and data storage is almost the same.

Moreover, packet loss evaluations are also carried out during the acquisition process. Regardless of the different acquisition duration, the MP and SW did not experience packet loss. It indicates that FIFO, queues, buffers, and other mechanisms for concurrency and speed mismatch in various parts of the BEATS work well and not cause the data loss or discarding. With no data loss and ultra-low delay, the stability and efficiency of the system can be verified.


\section{Application Cases}

The first two cases introduces the acquisition of EEG signals and other electrophysiological signals including ECG, electrooculogram (EOG), and EMG. The third case takes emotion elicitation experiment as an example to construct experiments and gives the data analysis results.

\subsection{Alpha Wave of EEG Acquisition}

\begin{figure}[ht]
	\centerline{\includegraphics[width=0.45\textwidth]{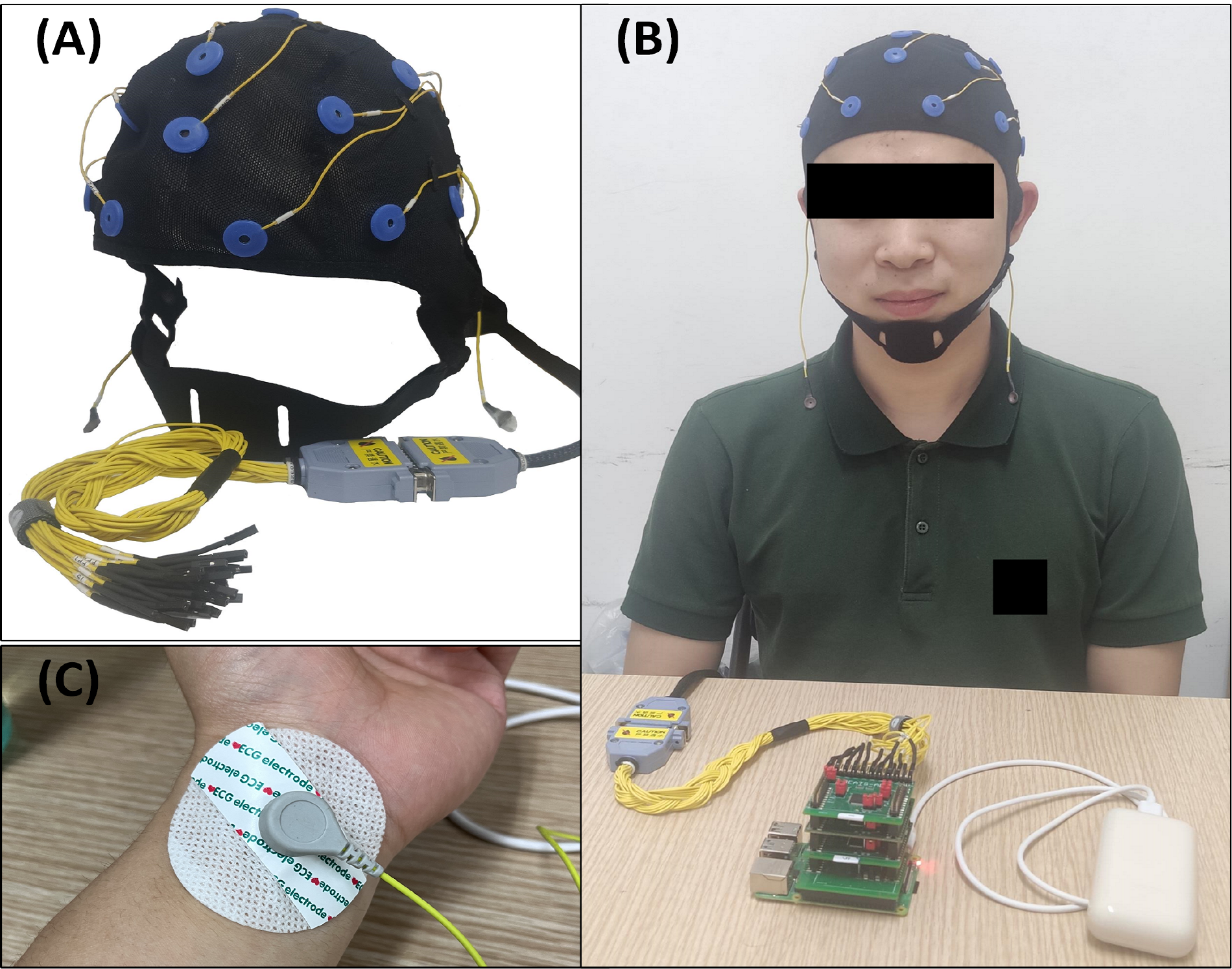}}
	\caption{(A) The used wet electrode EEG Cap. (B) The participant wearing the EEG cap with BEATS connected. (C) The patch electrode with a snap connector.}
	\label{fig:eegcap}
\end{figure}

For the EEG acquisition, the commercially available wet electrode EEG cap is used, as shown in Fig. \ref{fig:eegcap} (A). It has low impedance and better signal quality than other types of electrodes. The EEG cap used has 23 electrodes, two of which are reference and ground electrodes, two of which are additional reference electrodes, and the remaining 19 channels of which are EEG electrodes. Three ADC-boards and a motherboard are plugged together for EEG acquisition. The placement of electrodes is consistent with the standard 10-20 system. The photo of the participant wearing the EEG cap with BEATS connected is shown in Fig. \ref{fig:eegcap} (B).

\begin{figure}[ht]
	\centerline{\includegraphics[width=0.5\textwidth]{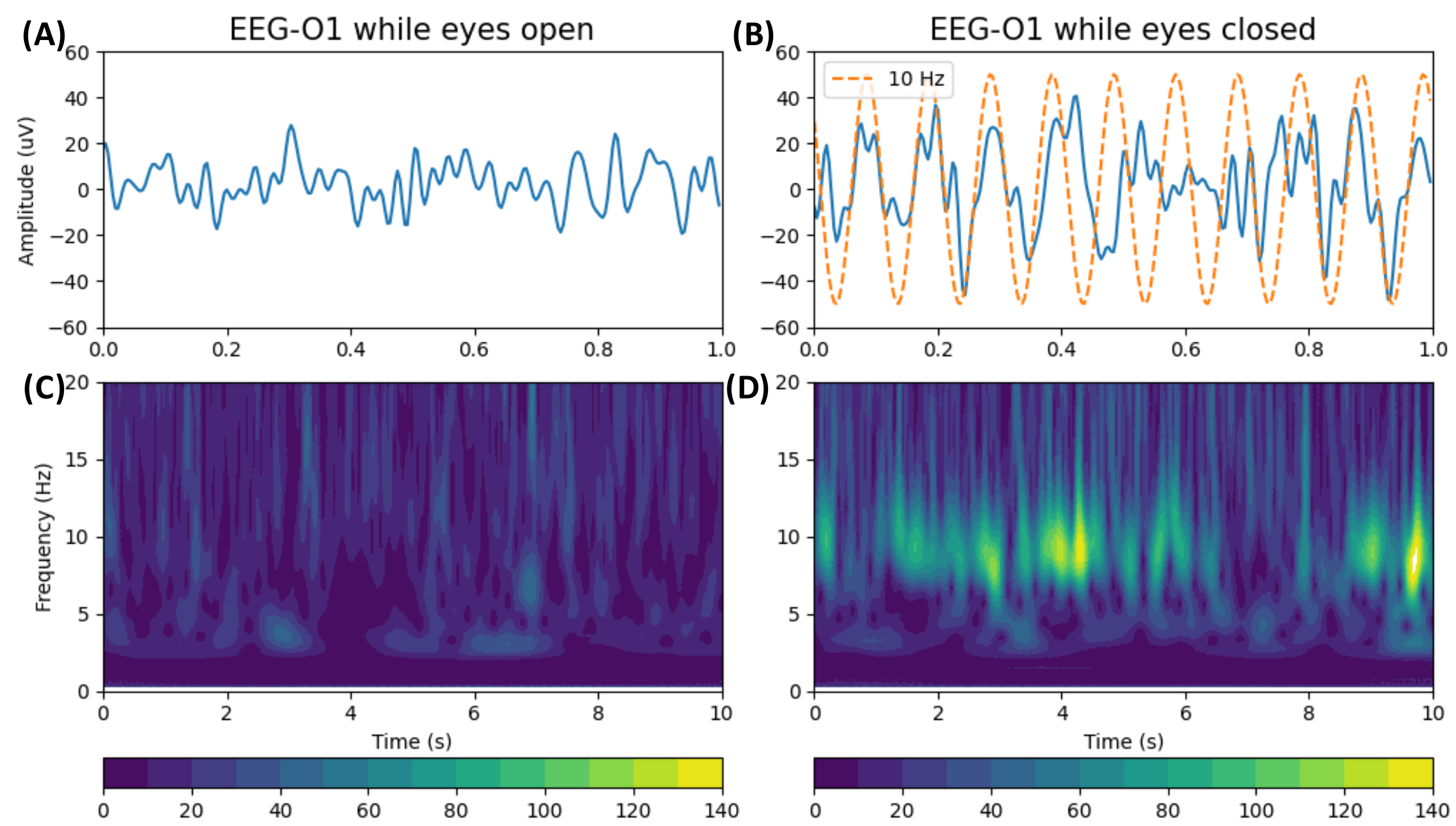}}
	\caption{(A) (B) The time-domain EEG signals in the O1 when the eyes are open and closed. (C) (D) The time-frequency spectrum when the eyes are open and closed.}
	\label{fig:alpha}
\end{figure}

EEG signals can be divided into different bands according to frequency ranges. Alpha wave (8-13 Hz) is one of the basic waves of EEG, which is most obvious in the occipital lobe and posterior parietal lobe. It disappears when eyes open and reappear when eyes close. Therefore, EEG signals when the eyes are opened and closed are collected to observe this phenomenon. Signals from the occipital region are used for analysis, as the alpha wave changes are most pronounced here. When the eyes are open and closed, 1-second time-domain waveforms are cut out for observation, as shown in Fig. \ref{fig:alpha} (A) (B). It can be seen that alpha waves reappears when eyes are closed. In addition, 10-second time-frequency spectrums of eyes open and closed are cut out for observation, as shown in Fig. \ref{fig:alpha} (C) (D). When the eyes are open, the activity of the alpha waves is not obvious. When the eyes are closed, the energy of the alpha wave is significantly higher than the other frequency bands.

\subsection{Multi-electrophysiological Signal Acquisition}

\begin{figure}[ht]
	\centerline{\includegraphics[width=0.5\textwidth]{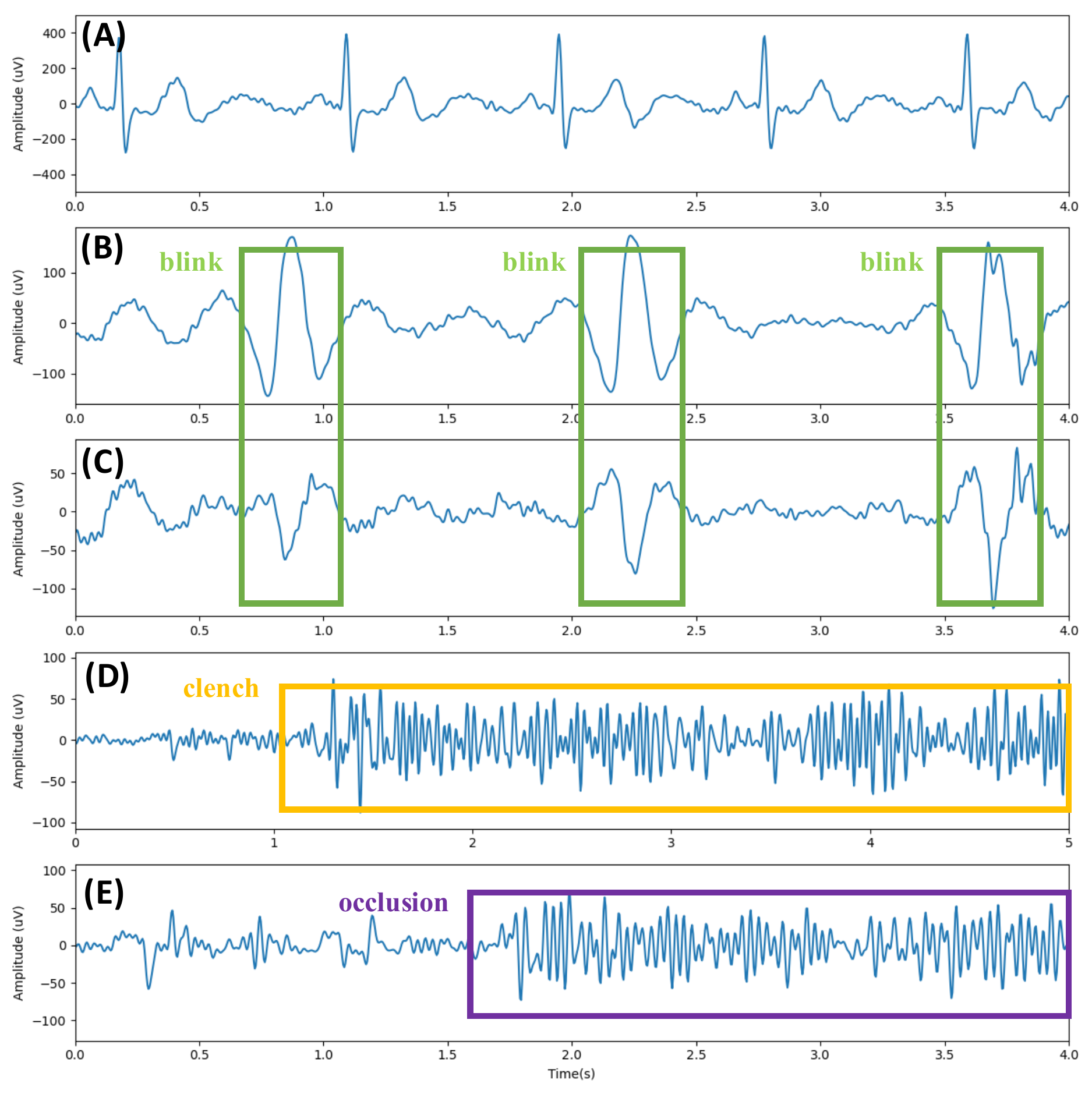}}
	\caption{The ECG (A), EOG (B) (C), and EMG (D) (E) signals acquired.}
	\label{fig:signals}
\end{figure}

Although BEATS is developed specifically for EEG acquisition, it has higher requirements compared to other electrophysiological signals \cite{b3}. For the ECG, EOG, and EMG signals acquisition, an 8-channel BEATS with an ADC-board and a motherboard is used, and the electrodes used are the patch electrodes with a snap connector, as shown in Fig. \ref{fig:eegcap} (C). Specifically, the ECG electrodes are placed on the wrist of the left hand. The EOG electrodes are placed at the upper and lower positions of the eyes. The EMG electrodes are connected to the right forearm at near positions and the mandibular muscles. The mastoid processes of the left and right ears are used as the reference electrode and the bias electrode respectively. The acquired signals are shown in Fig. \ref{fig:signals}.

Obvious and regular ECG signals can be seen in Fig. \ref{fig:signals} (A) and the PQRST wave in ECG waveform can also be clearly seen. Because the EOG signals are collected at the upper and lower positions of the eyes, the upper and lower EOG should be inverted at the same time. After intercepting a period of data before and after blinking, the inverse state of EOG can be observed in Fig. \ref{fig:signals} (B) (C). Since the reference electrode and the position of EMG acquisition cross the heart, there is a pulse signal in the original EMG signal. Using the two EMG signals as a reference to each other, the pulse signal can be offset, and the current position EMG signal can be obtained. Intercepting the EMG signal before and after clenching and occlusion, it can be seen that the amplitude and frequency of EMGs increase, as shown in Fig. \ref{fig:signals} (D) (E).

\subsection{Emotion Elicitation Experiment}
Emotion is an important human state. The emotions of the participants are elicited by watching video clips with specific emotions. After each video clip ended, the participants assess and give feedback on their own emotional state. The principles and procedures of this experiment follow the EEG datasets commonly used for emotion recognition, like DEAP \cite{b25} and SEED \cite{b26}, and it is the main paradigm in the current emotion-related research. During the experiment, the EEG signals are recorded in real time using BEATS. The acquisition scheme is the same as the EEG acquisition case mentioned before.

\subsubsection{Materials Preparation}

The video clips used to elicit emotions are derived from \cite{b27}, which provides a publicly available music video emotion dataset that integrates and augments the video clips in DEAP. From this dataset, video clips with positive, neutral, negative emotions are selected for subsequent experiments. Each video clip is approximately 30 seconds, and the clips with a length greater than 30 seconds are selected and trimmed to the same duration of 30 seconds.

In the process of selection, volunteers are asked to watch these clips and record the emotion type they were elicited. The evaluation process for each volunteer is completely independent and each clip is evaluated by at least 3 volunteers. Then, clips having a different label with volunteers' records are rejected, which indicates that these chips cannot elicit corresponding emotions. During the clips selection process, the number of clips for the three emotions is kept consistent. In the end, 30 clips of each emotion, a total of 90 clips, are selected for subsequent experiments.

\begin{table*}[t]
\centering
\caption{Comparison with existing boards}
\label{tab:cmp}
\begin{tabular}{rccccccccc}
\hline
\multicolumn{1}{c}{} &
  Channels &
  \begin{tabular}[c]{@{}c@{}}Sampling Rate\\ (Hz)\end{tabular} &
  Open Source &
  GUI &
  \begin{tabular}[c]{@{}c@{}}Weight\\ (g)\end{tabular} &
  Size (cm) &
  \begin{tabular}[c]{@{}c@{}}Cost $^{1}$\\ / (Price)\end{tabular} &
  \begin{tabular}[c]{@{}c@{}}Operating Time $^{2}$\\ (Power)\end{tabular} &
  Wireless \\ \hline
Open BCI Cyton                            & 16   & 250    & Yes & Yes & 260  & -                           & (\$950)      & 24 h          & RFDuino     \\
Emotive Epoc                              & 14   & 128    & No  & -   & 125  & $9 \times 15 \times 15$     & (\$799)      & 12 h          & Proprietary \\
BioSemi                                   & 256  & 2-16k  & No  & -   & 1.1k & $12 \times 15 \times 19$    & ($>$\$10000) & 5 h / wired   & No          \\
g.tec nautilus                            & 64   & 500    & No  & -   & 360  & $7.8 \times 6 \times 3.6$   & ($>$\$1000)  & 10 h          & Bluetooth   \\ \hline
Creamino \cite{b7}      & 32   & 500    & Yes & Yes & -    & -                           & \$170        & -             & No (USB)    \\
\cite{b8}             & 24   & 250    & No  & No  & 115  & -                           & \$225        & 24 h          & Wi-Fi       \\
\cite{b9}          & 64   & 1k     & No  & Yes & -    & $10 \times 5$               & \$116        & (250 mAh)     & Wi-Fi       \\
\cite{b10}                 & 32   & 2k     & No  & No  & -    & $3.4 \times 3 \times 1.5$   & -            & -             & Wi-Fi       \\
CochlEEG \cite{b11}  & 8    & 4k     & No  & Yes & 47   & $6.6 \times 3.3 \times 2.3$ & -            & -             & No (USB)    \\
\cite{b12}                    & 8    & 4k     & No  & Yes & 90   & $5.6 \times 5.1$            & -            & 5 h (200 mAh) & Bluetooth   \\ \hline
BEATS (ours)                              & 32   & 4k     & Yes & Yes & 352  & $9 \times 5.6 \times 8.2$   & \$400        & 24+ h / wired & Wi-Fi       \\ \hline
\multicolumn{10}{l}{$^{1}$ The upper part indicates the price of commercial products, and the lower part indicates the cost of self-designed products.} \\
\multicolumn{10}{l}{$^{2}$ wired: the system can be powered by a wired connection to the grid without the need for the battery.}                       
\end{tabular}
\end{table*}

\subsubsection{Experiment Procedures}
Healthy participants are selected to conduct the experiment, and they are all right-handed college students aged 18 to 25, including males and females. Each participant is asked to watch 15 selected clips continuously, which includes 5 clips of each emotion. Clips are randomly selected by label from all selected videos, and the order in which they are played is also random. Each time after watching the clip, there is a few seconds for the participant to evaluate the emotion state during watching the clip. Participants need to provide a feedback on the elicited emotion label and intensity on a scale of 0 to 10. Comparing the emotion labels reported by the participants and the expected labels, most of the elicited emotions are consistent with expectations. In the results of 120 times clips watching, there are 6 times of the elicited emotions are not as expected and these data are discarded.

\subsubsection{Spectral Analysis}
\begin{figure}[ht]
    \centerline{\includegraphics[width=0.48\textwidth]{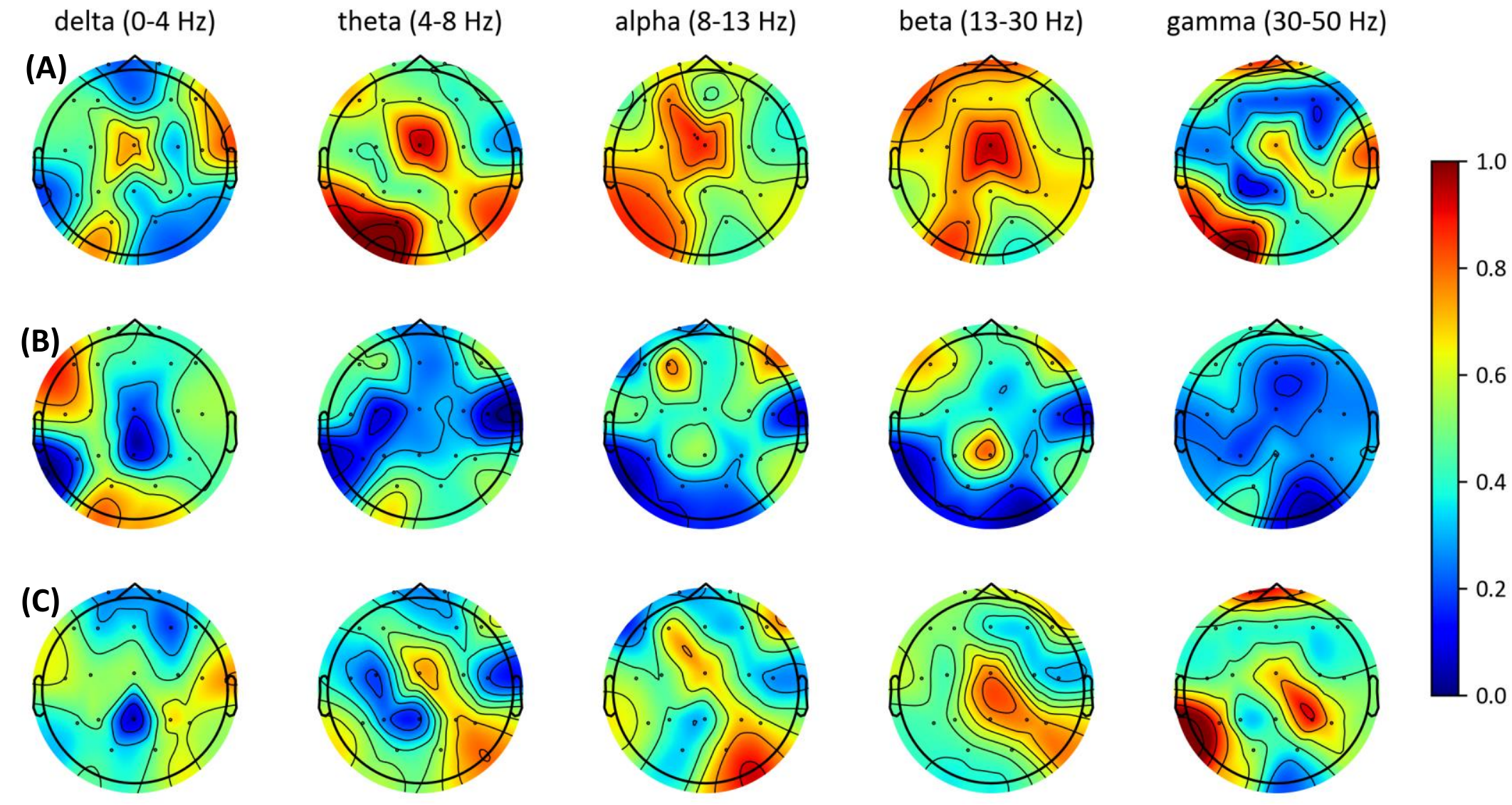}}
    \caption{The band power of five frequency bands in different brain regions. (A) Positive. (B) Neutral. (C) Negative.}
    \label{fig:bmap}
\end{figure}

First, the power spectral density (PSD) of the EEG signals in each clip is calculated.
Based on the five frequency bands of the EEG, the band power is calculated by accumulating and averaging the PSD.
Then, according to the electrode positions on the brain, the band powers are mapped to an EEG topographic map as shown in Fig. \ref{fig:bmap}.
For clips with the same label, their band power is added up.

For positive emotion, it shows that power of theta, alpha, and beta frequency bands is higher than others, especially in the beta band. For neutral emotion, its overall power is lower than both positive and negative emotions whereas the power in the delta band is slightly higher than others. Compared with the positive emotion, the negative emotion's power is lower in most bands but slightly higher in the gamma band. The negative emotion's power is concentrated in the right back of the brain, whereas the positive emotion's power is mainly concentrated in the middle and front of the brain.

\section{Conclusion}

This paper presents an open-source, high-precision, multi-channel EEG acquisition tool system named BEATS. It achieves similar features with other state-of-the-art implementations including cost-effectiveness, portability, etc. The comparison between different implementations is given in Table \ref{tab:cmp}.

In addition to similar features, BEATS mainly has the contributions as follows. First, BEATS implements 32-channel EEG acquisition at a sampling rate of 4 kHz. It has a higher channel number for the same sampling rate or a higher sampling rate for the same channel number and can support more channel numbers if a lower sampling rate is configured. Second, due to the concurrency mechanism including the interrupt, FIFO, DMA, thread, timer, the high precision, stability and efficiency of BEATS are guaranteed. After evaluation, the maximum delay is 0.07 s/h for 24 hours of continuous acquisition. Third, as an open-source system, BEATS provides complete design files and materials, which is capable of being quickly and easily reproduced.

Besides, BEATS implements a complete system from hardware to software, including PCBs using easy-to-access materials and a plug-in structure, a user-friendly GUI and a wireless data transmission, which make it can be easily set up. Meanwhile, it can also be able to acquire ECG, EOG, EMG, and other signals. Due to these exceptional properties of BEATS, it can be widely used in various research scenarios, and is ideal for long-term daily monitoring. In the future, we will further investigate the synchronization scheme of stimulus generation and signal acquisition to achieve higher precision.


\end{document}